\documentclass[fleqn,twoside]{article}
\usepackage{gc,cite}

\heads{Saibal Ray, Utpal Mukhopadhyay and Xin-He Meng}
      {Accelerating Universe with a dynamic cosmological term}

\begin{document}
\twocolumn[
\Arthead{13}{2007}{2 (50)}{142}{150}

\Title
{Accelerating Universe  with a dynamic cosmological term}

\Aunames{Saibal Ray\auth{1,a},
        Utpal Mukhopadhyay\auth{b} and Xin-He Meng\auth{c} }

\Addresses{
  \addr a {Department of Physics, Barasat Government College, Kolkata
        700 124, North 24 Parganas, West Bengal, India;\\
       IUCAA, Post Bag 4, Pune 411 007, India}
  \addr b {Satyabharati Vidyapith, Kolkata 700 126,
        North 24 Parganas, West Bengal, India}
  \addr c {Department of Physics, Nankai University,
        Tianjin 300 071, P. R. China;\\
       BK 21 Division of Advanced Research and Education in Physics,
        Hanyang University, Seoul 133-791, Korea}    }

\Rec{26 December 2006}
\Recfin{6 May 2007}

\Abstract
{Keeping in mind the current picture of an accelerating and flat Universe,
some specific dynamical models of the cosmological term $\Lambda$ have been
selected for investigating the nature of dark energy. Connecting the free
parameters of the models with the cosmic matter and vacuum energy density
parameters, it is shown that the models are equivalent. Using the selected
models, the present values of some of the physical parameters have been
estimated, and a glimpse at the past decelerating universe has also been
presented. It is observed that most of these cosmological parameters nicely
agree with the values suggested by the Type Ia Supernovae and other
experimental data.}

] %%%%%%%%%%%%%%
\email 1 {saibal@iucaa.ernet.in}

\section{Introduction}

The observations on supernova by the High-$z$ Supernova Search Team (HZT)
and the Supernova Cosmology Project (SCP) \cite{rie98,per98} have revealed
that, instead of slowing down, the expanding Universe is speeding up. An
intense search is going on, in both theory and observations, to unveil the
true nature of this acceleration. It is commonly believed by the
cosmological community that a kind of repulsive force which acts as
anti-gravity is responsible for gearing up the Universe some 7 billion
years ago. This hitherto unknown exotic physical entity is termed
as {\it dark energy\/}.

Now, there can be many variants of dark energy which can be responsible for
this accelerated universe, and variation in the forms of dark energy also
exhibit variation in expansion rates in different eras. So, there may be
more than one candidate which can be stamped as dark energy. For example,
one may select the so-called {\it cosmological constant\/}, introduced and
later abandoned by Einstein, as dark energy. But selection of the
cosmological constant as dark energy faces a serious fine-tuning problem
which demands that the value of $\Lambda$ must be 123 orders of magnitude
and 55 orders of magnitude larger on the Planck scale ($T \sim 10^{19}$ GeV)
and the electroweak scale ($T \sim 10^2$ GeV), respectively, than its
presently observed value. Moreover, the matter and radiation energy
densities of the expanding Universe fall off as $a^{-3}$ and $a^{-4}$,
respectively, where $a$ is the scale factor of the universe, while $\Lambda$
remains constant. This poses another disturbing fine-tuning problem. For
these two reasons, at present $\Lambda$ with a dynamical character is
preferred over a constant $\Lambda$, especially a time-dependent $\Lambda$
which has decreased slowly from its large initial value to reach its small
value at present \cite{ove98}. A scalar field $\phi$ with a potential
$V(\phi)$, which is known as {\it quintessence\/} and decreases slowly with
time, may be another candidate for dark energy.  Quintessence exerts
negative pressure and is dynamic in nature (recall that
$\Lambda_{\rm effective} = 8\pi G \rho_\phi$). However, in the present
article we have considered some phenomenological models of kinematical
$\Lambda$ which is assumed to be one of the dark energy candidates to
account for the accelerating expansion of the Universe.

Among the dynamical models of $\Lambda$ which are frequently used in the
literature, we have particularly presented here three types, viz.,
$\Lambda \sim (\dot a/a)^2$, $\Lambda \sim \ddot a/a$ and $\Lambda \sim
\rho$, where $a$ is the scale factor of the Robertson-Walker metric and
$\rho$ is the matter energy density. The first type of $\Lambda$-model was
proposed from dimensional arguments by Carvalho et al. \cite{car92} and
Waga \cite{wag93} and using another type of argument by Lima and Carvalho
\cite{lim94}, and it was subsequently taken up by several workers
\cite{sal93,arb94,wet95,arb97,pad01}. Using dimensional arguments,
Vishwakarma \cite{vis00} suggested the $\Lambda \sim \rho$ model, whereas the
second model mentioned above was dealt by Arbab \cite{arb03a,arb03b,arb04}
and Overduin and Cooperstock \cite{ove98}.

Now, a key to catch up the nature of dark energy lies in $w$, the equation
of state parameter which is nothing but the ratio of fluid pressure and
energy density of dark energy, viz., $w = p/\rho$. This parameter $w$ has
different forms in different models. In the present study, using the above
three forms of $\Lambda$, general solutions of the field equations are
obtained under the assumption that the Universe is flat. Also, particular
solutions, wherever needed, are discussed for the specific cases of matter-
and radiation-dominated universes related to three specific dynamic
cosmological terms. It is possible to show the equivalence of the three
models in terms of the solutions, obtained by connecting the free parameters
$\alpha$, $\beta$ and $\gamma$ of these models with $\Omega_m$ and $
\Omega_{\Lambda}$, the matter and vacuum energy density parameters of the
Universe, respectively. This will enable us to establish a relationship
between three parameters of the models in the pressureless dust and
electromagnetic radiation cases ($w=0,\ 1/3$).

In this connection, we would like to point out that, concerning the cases
$\Lambda \sim (\dot a/a)^2$ and $\Lambda \sim \rho$, it was already
mentioned by Vishwakarma \cite{vis2001cqg} that the estimates of the
parameters for flat models are the same. Therefore, in view of this, the
main purpose of the present paper is to reexamine the status of the
phenomenological approach of the dynamical $\Lambda$-term and to provide
more general result by including one more case $\Lambda \sim \ddot a/a$
into a systematic analysis. However, though there are innumerable
phenomenological $\Lambda$-decay laws available in the literature (see for
exhaustive lists \cite{ove98,sah00}), this particular case, viz. $\Lambda
\sim \ddot a/a$, is not included there. This case has been so far,
separately, taken up by Arbab \cite{arb03a,arb03b,arb04} and also by
Overduin and Cooperstock \cite{ove98} with a different approach. We have,
among other candidates of the list, purposely omitted the popular cases
like $\Lambda \sim t^{-2}$ and $\Lambda \sim a^{-2}$ since the first one
exactly coincides with that of the case $\Lambda \sim (\dot a/a)^2$ as $t
\sim H^{-1}$ where $H$ is the Hubble parameter, which is defined as $H =
\dot a/a$ and was extensively studied by several authors
\cite{ras72,end77,can77,lau85,ber86ncb,ber90ijtp,ber91prd,bee94,
al95grg,al96plb,lop96,vis2001cqg,arb03b,arb04}.
The second case, $\Lambda \sim a^{-2}$, which was first suggested through
dimensional arguments related to quantum cosmology by Chen and Wu
\cite{che90w} and also results from a contracted Ricci collineation along
the fluid-flow vector \cite{abd96,vis01grg}, is dropped here since this
case does not suit for our present scheme as will be clear from the field
equations of the next section.

Based on all the available observational information, some physical features
have been explored through the cosmological parameters, which are in good
agreement with the observationally obtained present data of the Universe.
These results are discussed in \sect 6 considering both the present
accelerating and the past decelerating Universe. Before this, we show the
ranges of the parameters $\alpha$, $\beta$ and $\gamma$ involved in
different $\Lambda$-models in \sect 5, whereas the equivalence of the
$\Lambda$-models is established in \sect 4. Sections 2 and 3 are related to
the Einstein field equations and their general solutions for different
$\Lambda$-dependent models.  In the concluding \sect 7, some discussion is
presented.

\section{Einstein's field equations}

Let us consider the Friedmann-Lema\^{\i}tre-Robertson-Walker (FLRW) metric
\bearr
    ds^2 = -dt^2 + a(t)^2\left[\frac{dr^2}{1 -
        kr^2} + r^2 (d\theta^2 + sin^2\theta d\phi^2)\right]
\nnn       %% 1
\ear
where the curvature constant $k = -1, 0, +1 $ for hyperbolic, flat and
closed models of the Universe, respectively.

The Einstein field equations are given by
\bear
  R^{ij} - \frac{1}{2}Rg^{ij} = -8\pi
        G\left[T^{ij} - \frac{\Lambda}{8\pi G}g^{ij}\right]
\ear           %% 2
where $\Lambda$ is the so-called {\it cosmological constant,\/} assumed here
to be time-dependent, viz., $\Lambda = \Lambda(t)$, and $c$, the velocity of
light in vacuum, is assumed to be unity (we thus use relativistic units).

For the spherically symmetric metric considered above, the Einstein field
equations with a time-dependent cosmological constant yield the following
two equations, called the Friedmann equation and the Raychaudhuri equation:
\bearr
    \left(\frac{\dot a}{a}\right)^2 + \frac{k}{a^2} =
            \frac{8\pi G}{3}\rho + \frac{\Lambda}{3},
\\ \lal
    \frac{\ddot a}{a} = - \frac{4\pi G}{3} (\rho +
                    3p)+ \frac{\Lambda}{3}.
\ear
 The energy conservation law can be written as
\bear
    8\pi G(p + \rho)\frac{\dot a}{a} =
        -\frac{8\pi G}{3}\dot \rho - \frac{\dot\Lambda}{3}.
\ear

Let us choose the barotropic equation of state
\bear
    p = w \rho              %% 6
\ear
 where the parameter $w$ can take the constant values $0$, $1/3$, $-1$ and
 $+1$ for dust, radiation, vacuum fluid and stiff fluid, respectively.

 Using \eq (6), \eq (4) transforms to
\bear                               %% 7
    \frac{\ddot a}{a} + \frac{4\pi G}{3} (1 + 3w)\rho =\frac{\Lambda}{3}.
\ear
 Differentiating \eq (3) with respect to the time coordinate $t$ and using
 \eqs (4)--(7) to eliminate $\rho$, we finally obtain the following
 equation:
\bearr
    \left(\frac{\dot a}{a}\right)^2 + \left[3\left(\frac{1+
    w}{1 + 3w}\right) - 1 \right]\frac{\ddot a}{a} + \frac{k}{a^2} =
        \left(\frac{1+ w}{1 + 3w}\right)\Lambda.
\nnn
\ear                                                 %% 8
 This is the dynamical equation relating the cosmic scale factor $a$ to a
 known value of the dynamic cosmological term $\Lambda$. It can readily be
 observed from the above equations (7) and (8) that $\Lambda$ depends on the
 factors $\ddot a/a$, $\rho$, $(\dot a/a)^2$ and $a^{-2}$ in a specific way.
 However, the inflation theory of the Universe predicts and the CMB detectors
 such as BOOMERANG \cite{bernardis00,bernardis02,netter02}, MAXIMA
 \cite{han00,lee01,bal01}, DASI \cite{hal02}, CBI \cite{sie03} and WMAP
 \cite{ben03,spe03} confirm that the Universe is spatially flat. Therefore,
 the $\Lambda \sim a^{-2}$ case, which is not suitable for $k=0$, is omitted
 here. However, for a detail study of this case, viz. $\Lambda \sim a^{-2}$,
 interested persons may consult the works done by Chen and Wu \cite{che90w},
 Abdussattar and Vishwakarma \cite{abd96} and Vishwakarma \cite{vis01grg},
 mentioned earlier, as well as \cite{oze86,abd92,cal92,car92,wag93,sil94,
 men96,abd97,jaf99}. We shall therefore consider the phenomenological models
 related to the cases $(\dot a/a)^2$, $\ddot a/a$ and $\rho$ only for
 $\Lambda$ and try to find solutions which will help us to model and
 explore the features of the Universe.

\section{Cosmological models for an accelerating Universe}

 If we use $\Lambda = 3\alpha (\dot a/a)^2 = 3\alpha H^2$, where $\alpha$
 is a constant and $H$ is the Hubble parameter, then for the flat universe
 ($k = 0$) \eq (8) reduces to
\bear
     2a \ddot a + (1 + 3w - 3w\alpha -3\alpha)\dot a^2 = 0.    %% 9
\ear
 Solving \eq (9), we get our general solution as
\bear                                                         %% 10
    a(t) \eql C_1t^{2/3(1 - \alpha)(1 + w)},
\\                                                               %% 11
    \rho(t) \eql \frac{1}{6\pi G(1 - \alpha)(1 + w)^2}t^{-2},
\\                                                                  %% 12
    \Lambda(t) \eql \frac{4\alpha}{3(1 - \alpha)^2(1 + w)^2}t^{-2},
\ear
 where $C_1$ is an integration constant.

 It is evident from \eqs (10)--(12) that $\alpha \neq 1$ for physical
 validity. Moreover, a repulsive $\Lambda$ demands positive $\alpha$ via
 \eq (12) while \eq (11) shows that, for positive $\rho$, the parameter
 $\alpha$ must be less than 1 and imposes the constraint $0 < \alpha < 1$.
 The case $\alpha \geq 1$ is either nonphysical or incompatible with a
 time-dependent $\Lambda$. This is because a solution with a variable
 $\Lambda$ is possible in the presence of matter only when $T^{ij}_{;j}
 \neq 0$ \cite{vis2002cqg}. Again, since we are dealing with a non-zero
 $\Lambda$, we have $\dot a \neq 0$. This means that, when $\Lambda \sim
 (\dot a/a)^2$, the expansion of the Universe never stops as long as
 $\Lambda \neq 0$.

 Similarly, if we set $\Lambda = \beta (\ddot a/a)$ and $\Lambda = 8\pi
 G\gamma \rho$, where $\beta$ and $\gamma$ are free parameters, then, for $k
 = 0$, it can be very easily shown that in both cases the scale factor
 follows the same type of power laws as in \eq (10) while, just as in
 \eqs (11) and (12), $\rho(t)$ and $\Lambda(t)$ are inversely proportional
 to $t$. It may be mentioned that for physical validity either $\beta<0$ or
 $\beta>3$ for the present model. On the other hand, for a non-negative,
 repulsive $\Lambda$ one needs to impose the condition $\gamma > 0$ while
 for positive $\rho$ it should be $\gamma>-1$. This means that $\gamma$
 is always a positive quantity.

\section{Equivalence of three forms of dynamic $\Lambda$}

 Now, let us explore the interrelations between $\alpha$, $\beta$ and
 $\gamma$ and hence the equivalence of different forms of the dynamic
 cosmological terms, viz., $\Lambda \sim (\dot a/a)^2$, $\Lambda \sim \ddot
 a/a$ and $\Lambda \sim \rho$.

 From \eq (10), differentiating it and then dividing by $a$, we get
\bear
    t = \frac{2}{3(1 - \alpha)(1 + w)H}
\ear
 where $H$ is the Hubble parameter, as mentioned earlier, and hence, for
 specific values of $\alpha$ and $w$, \eq (13) shows that $H \sim t^{-1}$.
 This point was indicated in the Introduction, and therefore we have omitted
 the case $\Lambda \sim t^{-2}$ from the present investigation.

 Using \eq (13) in (11) and the definition of the cosmic matter density
 parameter $\Omega_m(= 8\pi G \rho/3H^2)$, one gets
\bear
     \Omega_{m\alpha} = 1 - \alpha
\ear
 where $\Omega_{m\alpha}$ is the cosmic energy density parameter for the
 $\alpha$-related dynamic $\Lambda$-model.

 Again, using \eq (13) in (12) and the definition of the cosmic vacuum
 energy density parameter $\Omega_{\Lambda} = \Lambda/3H^2$, we have
\bear
        \Omega_{\Lambda\alpha}= \alpha,
\ear
 where, in a similar fashion, $\Omega_{\Lambda\alpha}$ is the vacuum energy
 density parameter for the $\alpha$-related dynamic $\Lambda$-model.

 Addition of \eqs (14) and (15) yields
\bear
    \Omega_{m\alpha} + \Omega_{\Lambda\alpha} = 1     %% 16
\ear
 which is the relation between the cosmic matter- and vacuum density
 parameters for a flat ($k=0$) universe.

 Equations similar to (16) can be obtained for $\beta$- and $\gamma$-related
 models as well. Thus, without loss of generality, we can write
\bearr                                                           %% 17
    \Omega_{m\alpha} = \Omega_{m\beta} = \Omega_{m\gamma} = \Omega_m,
\yyy
    \Omega_{\Lambda\alpha} = \Omega_{\Lambda\beta}           %% 18
        = \Omega_{\Lambda\gamma} = \Omega_{\Lambda},
\ear
 where $\Omega_m$ and $\Omega_{\Lambda}$ are the cosmic matter and vacuum
 density parameters. Therefore, in the absence of curvature, one can obtain
 the general relation
\bear
    \Omega = \Omega_{m} + \Omega_{\Lambda} = 1.         %% 19
\ear
 This analytical result is consistent with the observational constraint on
 the total energy density $\Omega$ of the Universe, where $\Omega =
 1.00^{+0.25}_{-0.30}$ due to the MAXIMA-I flight and COBE-DMR experiment
 \cite{bal01}, $\Omega = 1.05 \pm 0.08$ obtained from CBI-DMR observations
 \cite{sie03}, and $\Omega = 1.01 \pm 0.03$ ($68 \%$ CL) measured from the
 first acoustic peak in the angular power spectrum of CMB fluctuations
 \cite{reb03}.

 Now, \eqs (17) and (18) enable us to interrelate $\alpha$, $\beta$ and
 $\gamma$ with $\Omega_{m}$ and $\Omega_{\Lambda}$ as
\bear
    \alpha \eql \Omega_{\Lambda},
\\
 \beta \eql \frac{6\Omega_{\Lambda}}{2\Omega_{\Lambda} - \Omega_{m}(1 + 3w)},
\\
    \gamma \eql \frac{\Omega_{\Lambda}}{\Omega_{m}}.
\ear
 This result for the dust case of \eq (21) corresponds to Arbab's
 \cite{arb03b}. Thus, we find that the free parameter $\alpha$ here is
 nothing but the cosmic vacuum density parameter whereas $\gamma$ is
 the ratio of the cosmic vacuum and matter density parameters which,
 by virtue of \eq (19), provides
\bear
    \Omega = (1+ \gamma)\Omega_{m} = 1,
\ear
 which is another relation for the total cosmic energy density in the
 case of a flat universe.

 All the above general relations for $\alpha$, $\beta$ and $\gamma$ in terms
 of $\Omega_{m}$ and $\Omega_{\Lambda}$ also hold for the particular cases
 of dust ($w=0$) and radiation ($w=1/3$). It is interesting to note that,
 while the relations of $\alpha$ and $\gamma$ with the cosmic matter and
 vacuum density parameters are independent of $w$, the relations of $\beta$
 with $\Omega_{m}$ and $ \Omega_{\Lambda}$ are $w$-dependent.

 It can easily be shown that the particular solutions of $\Lambda \sim (\dot
 a/a)^2$ model for dust and radiation cases become identical with their
 corresponding counterparts for the other models in terms of the time
 dependences of $a$, $\rho$ and $\Lambda$ when expressed in terms of
 $\Omega_{m}$ and $ \Omega_{\Lambda}$. Therefore, these results imply that
 in $\Omega_{m}$ and $\Omega_{\Lambda}$ there are no distinctive features
 which could distinguish between the different forms of dynamic cosmological
 models, viz., $\Lambda \sim (\dot a/a)^2$, $\Lambda \sim \ddot a/a$ and
 $\Lambda \sim \rho$. Thus, starting from any of our $\Lambda$-models,
 since they are equivalent, we can arrive at the other relations.

 Now, from \eqs (20)--(22), we find that the parameters involved in the
 three dynamical relations are connected by
\bear
    \alpha = \frac{\beta(1 + 3w)}{3(\beta w + \beta - 2)}
            = \frac{\gamma}{1 + \gamma}.            %% 24
\ear
 This again shows that the three forms $\Lambda = 3\alpha (\dot a/a)^2$,
 $\Lambda = \beta (\ddot a/a)$ and $\Lambda = 8\pi G \gamma \rho$ are
 equivalent, and the three parameters $\alpha$, $\beta$ and $\gamma$ are
 connected by the relation (24). Thus it is possible to find out the
 identical physical features of others if any of the phenomenological
 $\Lambda$ relations is known. It can easily be seen that, for the dust case
 ($w=0$), \eq (24) relates $\alpha$ and $\beta$ as
\bear
    \alpha = \frac{\beta}{3(\beta - 2)},                %% 25
\ear
 which is Arbab's result\cite{arb04}. Moreover, it can be observed that our
 $\gamma$ is identical to Majernik's $\kappa$ \cite{maj01,maj03}, where
\bear                                                        %% 26
     \kappa = \frac{1}{\Omega_{m}} - 1 =\frac{\Omega_{\Lambda}}{\Omega_{m}}
\ear
 for the present situation in view of \eq (22). He has also shown that this
 result, \eq (26), is derivable from an {\it ansatz\/} by which $\Lambda$
 is proportional to the stress-energy scalar $T=T^i_j$, the trace of the
 stress-energy tensor of ordinary matter $T^j_i$, and is Lorentz-invariant.
 In this regard, following Majernik \cite{maj01}, it can be mentioned here
 that determination of the parameter $\gamma$ entirely depends on the cosmic
 matter density parameter or both the matter and vacuum density parameters.
 Thus this relation constrains the value of $\gamma$ and will be discussed
 in the next section.

 Another point to be mentioned here that \eq (19) and hence (23), via (22),
 is nothing but another form of the Friedmann equation (3) for the flat
 Universe. Thus, it is interesting to note that \eq (26) also represents the
 Friedmann equation. Therefore, starting from any of our $\Lambda$-models,
 since they are equivalent, we can arrive at the Friedmann field equation
 without any special assumption.

\section{Ranges of the parameters $\alpha$, $\beta$, $\gamma$}

 Recent measurements have given a wide range of values for $\Omega_{m0}$ and
 $\Omega_{\Lambda0}$. The first flight of the MAXIMA balloon-borne
 experiment (MAXIMA-I) combined with COBE-DMR resulted in $0.25 <
 \Omega_{m0} < 0.50$ and $0.45 < \Omega_{\Lambda0} < 0.75$ \cite{bal01}.
 Observations of SNeIa combined with the total energy density constraints
 from CMB \cite{reb03} and combined gravitational lens and stellar dynamical
 analysis \cite{koo03} lead to $\Omega_{m0} \sim 0.3$ and $\Omega_{\Lambda0}
 \sim 0.7$. The pinpoint values of these parameters as obtained by Sievers
 et al. \cite{sie03} and Spergel et al. \cite{spe06} are
 [$\Omega_{m0}$, $\Omega_{\Lambda0}$]
    = [$0.34 \pm 0.12 $,$0.67^{+0.10}_{-0.13}$] and
 [$\Omega_{m0}$, $\Omega_{\Lambda0}$]=
    [$0.249^{+0.024}_{-0.031}$, $0.719^{+0.021}_{ -0.029}$],
    respectively. These and other results are listed in Table 1.

\begin{table}
\caption{Values of $\Omega_{m0}$ and $\Omega_{\Lambda 0}$ from
various observational sources} \label{tab1}

{\small
 \tabcolsep 6pt

 \medskip
\begin{tabular}{@{}llrrrrlrlr@{}}
\hline \\[-9pt]
Source \& Reference & Year & $\Omega_{m0}$ & $\Omega_{\Lambda 0}$\\
\hline \\[-6pt]
Efstathiou et al. \cite{efs98}        &1998
&$0.25^{+0.18}_{-0.12}$&$0.63^{+0.17}_{-0.23}$\\
    \ \ (SNeIa + CMB) & & &                     \\
Riess et al. \cite{rie98} &1998 &$0.24^{+0.56}_{-0.24}$
    &$0.72^{+0.72}_{-0.48}$\\
    \ \ (SNeIa + MLCS) &  &       &                     \\
Perlmutter et al. \cite{per99}&1999&$0.4 \pm 0.1$       &0.7  \\
    \ \ (SNeIa + LSS) & & & \\
Balbi et al. \cite{bal01}   &2001 &$0.25{-}0.50$ & $0.45-0.75$\\
    \ \ (MAXIMA-I${+}$COBE)        &       & &  \\
Rebolo \cite{reb03} &2003   &0.30        &0.70     \\
    \ \ (SNeIa + CMB)    && &                     \\
Koopmans et al. \cite{koo03} &2003 & 0.30 &0.70 \\
    \ \ (Lens + SD) &       & & \\
Sievers et al. \cite{sie03}  &2003   &$0.34\pm 0.12$
  &  $0.67^{+0.10}_{-0.13}$\\
    \ \ (CBI)                    &       &&           \\
Barris et al. \cite{bar04}            &2004 &0.33 & 0.67 \\
    \ \ (IfA Deep Survey)        &       &            &            \\
Astier et al. \cite{ast06}  &2006 &$0.31 \pm 0.21$ & $0.80 \pm 0.31$ \\
    \ \ (SNLS) & & &\\
Spergel et al. \cite{spe06} &2006 &$0.249^{+0.024}_{-0.031}$ &
    $0.719^{+0.021}_{ -0.029}$ \\
    \ \ (WMAP + SNLS) & &  &  \\
 \hline
 \end{tabular}       }
\end{table}

 Considering the values in Table 1, we particularly prefer the matter density
 parameter as $\Omega_{m0} = 0.330 \pm 0.035$
 \cite{vis2002cqg,turner02,reb03,alc04}. This gives us an opportunity to
 obtain ranges of $\alpha_0$, $\beta_0$ and $\gamma_0$ (the  values of
 $\alpha$, $\beta$ and $\gamma$ at the present epoch) which can, using
 \eq (24), be obtained as $0.635 \leq \alpha_0 \leq  0.705$, $3.417
 \leq \beta_0 \leq 4.674$ and $1.739 \leq \gamma_0 \leq 2.389$ in the dust
 case. Thus we find that using our models we are able to obtain the range of
 $\beta$ smaller than Arbab's \cite{arb04} which was $3 < \beta < 4.5$ for
 dust.

 Again, if we recall the quintessence equation of state $p_Q = w_Q\rho_Q$
 where $w_Q = -\Omega_{\Lambda}$, we can easily obtain the relations between
 $\beta$ and $w_Q$ as $\beta = 6w_Q/(1 + 3w_Q)$ for $w=0$. Using the above
 range of $\beta_0$, we can calculate the range of $w_Q$ in dust as $-0.705
 \leq w_Q \leq -0.635$. It is interesting to note that the above result is
 consistent with the accepted range of $w_Q$ which is $-1<w_Q<0$. However,
 in the present investigation we are not concerned with the quintessence
 case and show the range of $w_Q$ as a check only.

\section{Features of the models}

\subsection{Physical parameters at the present accelerating epoch}

Now, a search for the status of $\Lambda$ rests on some observational
results from high-redshift type Ia Supernovae (SNeIa), the cosmic microwave
background radiation (CMBR) and other sources which inform us that the
present Universe is composed of about 30\% of ordinary matter and 70\% of
dark energy. Thus, in the present Universe, the vacuum density parameter
$\Omega_{\Lambda 0}$ is dominant over the matter density parameter
$\Omega_{m0}$. Determination of the Hubble parameter, a measure of the rate
of cosmic expansion, has been done by several authors based on different
values of density parameters as shown in Table 1. However, it is to be
noted that there exists a certain amount of uncertainty in the value of
$H_0$, as is obvious from Table 2.

\begin{table}
\caption{Values of $H_0$ and $t_0$ from various observational sources}
\label{tab2}

{\small
 \tabcolsep 5pt
\def\yy{\\[3pt]}

\medskip
\begin{tabular}{@{}llrrrrlrlr@{}}
\hline  \\[-9pt]
    Source \& Reference & Year & $H_0$\cm & $t_0\quad\ $ \\
      &    & km/(s$\cdot$ Mpc) & Gyr\ \ \ \\
\hline \\[-6pt]
Sandage, Tammann \cite{san97}$\nhq$          &1997 &$55 \pm
5$ &$13.5 \pm 2.5$\yy Sandage et al. \cite{san98} &1998 &
&$13.5^{+2}_{-3}$
\yy   Parodi et al. \cite{par00} &2000 &$58 \pm 6$ &-        \yy
  Birkinshaw \cite{bir99} &1999 &$60 \pm 10$ &-
\yy
  Saha et al. \cite{saha99} &1999 &$60 \pm 2$&-\yy   Jha
et al. \cite{jha99} &1999 &$64^{+8}_{-6}$&$14.1 \pm 1.6$
\yy   Perlmutter et al. \cite{per98} & 1998 &65              &14.9
\yy
  Riess et al \cite{rie98} &1998   &$65.2 \pm 1.3$ &$14.2 \pm
1.7$\yy   Tonry et al. \cite{ton03} &2003 &$58.8-72.3$&$13.1$
\yy   Sievers et al. \cite{sie03} &2003 &$66 \pm 11$  &$14.2 \pm
1.3$\yy   Tegmark et al. \cite{teg03} &2003
&$66^{+6.7}_{-6.4}$&$14.1^{+1.0}_{-0.9}$\yy
Rebolo \cite{reb03} &2003 &$60-75$      &$13.0 \pm 1.5$ \yy
Altavilla et al. \cite{alt04}] &2004 &$(68{-}74) \pm 7 $&-   \yy
  Cardone et al. \cite{car04}&2004 &$56{-}88$ &$13.1{-}14.3$\yy
 Freedman et al. \cite{fre01}
&2001   &$72 \pm 8$& $13$\yy Knox et al. \cite{kno01} &2001 &
&$14.0 \pm 0.5$\yy Ferreras et al. \cite{fer01} &2001 &
&$13.2^{+1.2}_{-0.8}$\yy Alcaniz \cite{alc04} &2004   & &$13.7 \pm
0.2$
\yy
  Freedman, Turner \cite{fre04} &2004   &$72 \pm 7$ &$13.0
\pm 1.5$\yy   Spergel et al. \cite{spe03} &2003   &$72 \pm 5$
&$13.4 \pm 0.3$
\yy
  Spergel et al. \cite{spe06} &2006 &$73.4^{+2.8}_{-3.8}$
&$13.73^{+0.13}_{-0.17}$\yy
 Freedman \cite{fre96} &1996 &$73 \pm 6$&-  \yy
Riess et al. \cite{rie05} &2005 &$73 \pm 9$ &-\yy
Peebles, Ratra \cite{pee03} &2003   &$73 \pm 10$&-   \yy
Blakeslee et al. \cite{bla99} &1999 &$74 \pm 4$&-\yy
Koopmans et al. \cite{koo03} &2003 &$75^{+7}_{-6}$&-\yy
Efstathiou \cite{efs95} &1995 & $80$ &-
\yy
Jacoby et al. \cite{jac92} &1992 &$80 \pm 11$&-\yy Freedman et
al. \cite{fre94} &1994   &  $80 \pm 17$ &-
\yy
Pierce et al. \cite{pie94}                &1994 & $87 \pm 7$ &-
\\ \hline
\end{tabular}
}
\end{table}

The data of Table 2 indicate that the present value of the Hubble parameter
is, in general, centralized at $72 \pm 8$ kms$^{-1}$Mpc$^{-1}$. Even the
most recent value ($73.4^{+2.8}_{-3.8}$) as obtained from WMAP by Spergel et
al. \cite{spe06} lies well within this range. Assuming this value of $H_0$
and $\Omega_{m0} = 0.330 \pm 0.035$ \cite{vis2002cqg,turner02,reb03,alc04},
the present age ($t_0$), the present matter density ($\rho_0$), the present
value of the cosmological term ($\Lambda_0$) and the value of the
deceleration parameter at the present era ($q_0$) have been calculated using
our equivalent models. All values of $\rho_0$ and $q_0$ are in nice
agreement with the modern concept of an open, accelerating universe.
Moreover, the values of $\Lambda_0$ support the idea of a small non-zero
cosmological parameter which is slowly decreasing in time, and at present
$\Lambda_0$ lies within $1 \times 10^{-35}$ s$^{-2}$ --- $2 \times 10^{-35}$
s$^{-2}$, which agrees with the results of Carmeli \cite{carmeli02} and
Carmeli and Kuzmenko \cite{car02k}, where they obtain the value of
$1.934 \times 10^{-35}$ s$^{-2}$. All values of $\rho_0$ are one order of
magnitude smaller than $10^{-29}$ g/cm$^3$, the critical density of the
Universe. For the matter-dominated case, various results can be obtained for
$t_0$ by finding $H_0$ for different $\Omega_{m0}$ (see Table 3 for detail).

\begin{table}
\caption{Age of the Universe from the present models} \label{tab3}
\centering
\bigskip
{\small

\begin{tabular}{@{}lcc@{}}
\hline \\[-9pt]
       $\Omega_{m0}$  &$H_0$            &$t_0$\\
               & kms/(s$\cdot$Mpc)      & Gyr \\
\hline \\[-9pt]
       0.33           &64               &27.79\\
               &72                      &24.70\\
               &80                      &22.23\\
\hline \\[-9pt]
       0.365          &64               &25.13\\
               &72                      &22.33\\
               &80                      &20.10\\
\hline \\[-9pt]
       0.40           &64               &22.93\\
               &72                      &20.38\\
               &80                      &18.34\\
\hline \\[-9pt]
       0.46           &64               &19.94\\
               &72                      &17.72\\
               &80                      &15.95\\
\hline
\end{tabular}   }
\end{table}

It is clear from Table 3 that, for the lower value of $\Omega_{m0}$, the age
becomes very high, whereas a higher value of the matter density parameter,
say, $\Omega_{m0} = 0.46$ provides a more realistic result for the age of
the Universe with gradual increase of the Hubble parameter. The best result
is therefore obtained for $\Omega_{m0} = 0.46$ (the upper limit of Sievers
et al. \cite{sie03}) and $H_0 = 80$ kms$^{-1}$Mpc$^{-1}$ (the upper limit
of Refs.\,\cite{kno01,fer01,ton03,alc04}) is $15.95$ Gyr. This result
exactly coincides with the upper limit of the value of Riess et al.
\cite{rie98} which is $14.2 \pm 1.7$ Gyr as obtained from SNeIa observations
and also very close to the values obtained by Sievers et al. \cite{sie03}
and Tegmark et al. \cite{teg03} as predicted by the WMAP data and CMB
observations.

\subsection{Physical parameters in the past decelerating period}

Using \eq (10), one can obtain an expression for the
deceleration parameter $q$ as
\bear                                                     %% 27
     q = - \frac{a\ddot a}{\dot a^2}= \frac{3(1 - \alpha)(1 + w)}{2} - 1.
\ear
Thus, for an accelerating universe
\bear
    \alpha >  \frac{1 + 3w}{3(1 + w)}.              %% 28
\ear
From \eq (28), it is evident that for dust ($w=0$) an accelerating universe
requires $\alpha > 1/3$. Now, $\alpha$ being the cosmic vacuum density
parameter by virtue of \eq (20), we find that our model fits an accelerating
universe since the modern accepted value of $\Omega_{\Lambda 0}$ is about
0.7 (Refs.\,\cite{tur02hubble,fra03,per03,car04} and see also Table 1) and
is much larger than $1/3$. Thus \eq (19) implies that the value of
$\Omega_{m}$ is 0.3, which provides $q_0 = - 0.50 \pm 0.05$ for the dust
case which can nicely accommodate the currently accepted value related to
the accelerating Universe \cite{tri97,efs98,sah99}. Again, $q$ will be
positive if $\alpha$ is less than $0.3$. Thus, for a decelerating universe,
the cosmic vacuum density parameter should be smaller than 0.3, which
is also consistent with the modern ideas. Therefore, we find that, within
our models, one can investigate accelerating as well as decelerating phases
of the cosmic expansion since $q$ depends on $\alpha$.

Now, it has already been mentioned that the expanding Universe, which is
about 14 Gyr old, entered into the present accelerating phase about 7 Gyr
ago. Therefore, for about 8 Gyr earlier, i.e.,  when the Universe was about
6 Gyr old, it was passing through a period of deceleration. Let us try to
estimate the values of some of the physical parameters when the age of the
Universe was 6 Gyr. From \eq (27) it is easy to obtain $q =(1.5 \Omega_{m} -
1)$ for $w=0$. We have already seen that $q$ will be positive for $\Omega_m
> 0.66$. Putting $\Omega_m = 0.67$, we find that $q>0$, i.e. the Universe
was indeed in a decelerating phase. Assuming $\Omega_m=0.67$ and $t=6$ Gyr,
we can estimate the values of $H$, $\rho$ and $\Lambda$, which are given by
$H \sim 179$ km\ s$^{-1}$Mpc$^{-1}$, $\rho= 3.3 \times 10^{-29}$ g cm$^{-3}$
and $\Lambda= 2.74 \times 10^{-35}$ s$^{-2}$, respectively. Similarly, for
the radiation case, $\Omega_m$ and $q$ are related by $q= (2\Omega_m -1)$.
Thus $q$ will be positive for $\Omega_m > 0.5$. If we assume $\Omega_m =
0.6$ and $t= 6$ Gyr, then the estimated values of $H$, $\rho$ and $\Lambda$
are 135 km\ s$^{-1}$Mpc$^{-1}$, $2.08 \times 10^{-29}$ g cm$^{-3}$ and $2.33
\times 10^{-35}$ s$^{-2}$, respectively. We thus find that in both cases
($w= 0, 1/3$) $\rho$ was above the critical density, which means that the
expanding Universe with a decelerating mode had a closed geometry.  The
value of the Hubble parameter, a measure of the expansion rate of the
Universe was slower in the radiation era than that of matter-dominated era.
Also, we find that the value of $\Lambda$ was slightly above its present
value, which justifies the idea of a dynamic $\Lambda$ which decreases very
slowly with time. Finally, assuming the present value of the Hubble
parameter as 72 km s$^{-1}$Mpc$^{-1}$, we see that the rate of decrease of
$H$ is about 13 km s$^{-1}$Mpc$^{-1}$Gyr$^{-1}$ for $w= 0$ and about $8$
km\ s$^{-1}$Mpc$^{-1}$Gyr$^{-1}$ for $w= 1/3$.

\section{Discussion}

In the present investigation, choosing some specific forms of dynamical
$\Lambda$, we were able to show the equivalence of those forms in terms of
the solutions obtained. While Arbab \cite{arb03b} has shown the equivalence
of the same three $\Lambda$-models in the context of a built-in cosmological
constant of Rastall \cite{ras72} and Al-Rawaf-Taha \cite{al95grg} type
models of modified general relativity, we have shown the equivalence of the
models with respect to their characteristic solutions in the framework of
Einstein's general relativity. It has already been mentioned that $\Lambda
\sim H^2$ and $\Lambda \sim t^{-2}$ are identical. In this context, it is
interesting to note that since $\ddot a/a$ is equal to $\dot H + H^2$ which
is again $ \sim t^{-2}$, then $\Lambda \sim \ddot a/a$ is also identical
with the above-mentioned cases. This is reflected in our solution sets via
\eq (24). Moreover, since $\ddot a/a = \dot H + H^2$, the $\Lambda \sim
\ddot a/a$ model can be thought of as a combination of two models, viz.,
$\Lambda \sim \dot H$ and $\Lambda \sim H^2$. Thus the $\Lambda \sim \ddot
a/a$ and $\Lambda \sim {{\dot a}^2}/a$ models become identical when $\dot H
= 0$. Now, $\dot H = 0$ implies a constant $H$, which in turn implies
exponential expansion and hence an inflationary scenario. Thus the idea of
inflation is inherent in the phenomenological model $\Lambda \sim \ddot
a/a$. Moreover, the $\Lambda \sim {{\dot a}^2}/a$ and $\Lambda \sim \ddot
a/a$ models cannot exist as separate entities during inflation.

We have also established a relation between $\alpha$, $\beta$ and $\gamma$,
the three parameters of the three forms of $\Lambda$ which ultimately yields
$\Omega_{m} + \Omega_{\Lambda}= 1$, the relation between the cosmic matter
and vacuum density parameters for a flat universe. It can be shown that this
particular relation between the density parameters also holds in the
radiation case. On the other hand, since $\Omega_{\Lambda} = \rho_{\Lambda}
/\rho_c$ and $\Omega_{m} = \rho_{m}/\rho_c$, it is clear from \eqs (20),
(21) and (22) that $\beta$ and $\gamma$ are independent of $\rho_c$, the
critical density of the Universe, whereas $\alpha$ depends on the critical
density. It can also be shown that, while $\beta$ and $\gamma$ decrease with
the age of the Universe, $\alpha$ increases as time passes.

Moreover, the present models represent a flat, accelerating Universe and do
not suffer from the low-age problem like many FRW models. Also, since the
present Universe is dark energy dominated, and the closest approximation to $t_0
\sim 20.10$ Gyr for the matter-dominated case can be obtained for the
specific choice of $\Omega_{m0}= 0.330 + 0.035$ and $H_0 = 72 + 8$ km\
s$^{-1}$Mpc$^{-1}$, our models point at the upper accepted limits of
$\Omega_{m0}$ and $H_0$. As has been shown earlier in Table 2 that there
is a certain amount of uncertainty in the value of $H_0$, the lower bound
being 50 km\ s$^{-1}$Mpc$^{-1}$ (see also Table 2 of Ref.\,\cite{san98} for
some more cases of a lower bound), whereas the upper bound is 97 km s$^{-1}$
Mpc$^{-1}$. The related values of the age of the Universe for our models
with these two extreme Hubble parameters are 25.52 Gyr and 13.15 Gyr,
respectively, when the matter density parameter is 0.46. Therefore, our
calculated value of $t_0$, which seems to be a bit over-aged and also
favours a $\Lambda$-dominated universe, can be accommodated to the accepted
age of the Universe within the error bar. Even though the values of $t_0$ in
case $w = 0$ show an excess with respect to the presently accepted age of
the Universe but, anyway, there is no low-age problem. Indeed, these values
are much higher than the age of the globular clusters which is $12.5 \pm
1.2$ Gyr \cite{martel98,gne01,cay01,cha02,gra03,marchi04}. In this
connection, we can also note that examples of higher age are not unavailable
in the contemporary literature
\cite{cow97,sta00,vis02mnras,alam02s,kal02,vis2002cqg,kal03l,alam03al}.
For example, Vishwakarma \cite{vis2002cqg} obtained, for $\Omega_{m0}= 0.330
\pm 0.035$ and $H_0 = 72 \pm 7$ km\ s$^{-1}$Mpc$^{-1}$, a remarkably
high age of the Universe, $t_0 \approx 27.4 \pm 5.6$ Gyr! But it is evident
from Table 2 that, whatever be the values of the Hubble parameter, the
experimental results for the age of the Universe lie around 14 Gyr. For
the present phenomenological models (including Vishwakarma's case
\cite{vis2002cqg}), the age of the Universe is inversely proportional to the
Hubble parameter. This provides a reasonable age of the Universe only for a
higher value of the Hubble parameter, which is also clear from the above
discussion. This is obviously a drawback of the present models unless a
higher value of $H_0$ is observationally established in the future.

\enlargethispage{-12pt}
In this regard, we would like to discuss the causal connection of our
models. We know that the proper distance $L(t)$ to the horizon is given by
\[
    L(t) = a(t) \int_{0}^{t} d\tau/a(\tau),
\]
and if the integral diverges, the model is causally connected. Since, for
$\Lambda \sim (\dot a/a)^2$, the scale factor $a(t)$ is given by \eq (10),
the proper distance $L(t)$ diverges if $1/3 < \alpha < 1$ or, in other
words, if $1/3 < \Omega_{\Lambda} < 1$ for the dust case and $\alpha<0.5$,
i.e., $\Omega_{\Lambda} < 0.5$ for radiation. Since the present Universe is
matter-dominated and the observational results indicate that at present
$\Omega_{\Lambda}\sim 0.7$, the Universe is causally connected in our
$\Lambda \sim (\dot a/a)^2$ model. Besides, since it has already been shown
that the present three phenomenological $\Lambda$-models are equivalent,
the causal connection of the Universe indicated in the above model implies
that the other models are also causally connected.

It should be mentioned that Arbab \cite{arb03b} has put his models to the
neoclassical tests like luminosity distance, angular diameter distance and
gravitational lensing, whereas we have tested the viability of our models
through age determination and some other measurements. So, in that respect
our work can be thought of as complementary to Arbab's investigation
\cite{arb03b}. Perspectively, we are studying new forms of a dynamic
cosmological term, such as those from the renormalization group, and its
confrontation with astrophysical observational data sets \cite{wan05,ren06},
expectating that this global description can help one to better understand
the mysterious dark energy nature and to alleviate the long-standing
cosmological constant problem.

Finally, it is to be noted here that, in general, $w$ is a function of time
\cite{che00,zhu01,pee03}. But the current observational data can hardly
distinguish between time-varying and constant equations of state
\cite{kuj02,bar05}, as demonstrated in some works (\cite{bar05} and
references therein). For this reason, $w$ is usually assigned a constant
value, as has been done by Caldwell et al. \cite{cal98} while dealing with
a relation between scalar field models and the XCDM parameterization.
Likewise, in the present work related to phenomenological $\Lambda$ models,
without showing a complete time evolution of $w$ (which is no doubt a
better representation), some specific cases are highlighted corresponding to
constant $w$. A more accurate analysis may be made in a later work by
considering $w=w(t)$.

\section*{Appendix}
\def\theequation{A.\arabic{equation}}
\sequ 0

  A comparative analysis of models
\bear
    \Lambda= \alpha(\dot a/a)^2, \cm \Lambda= \beta(\ddot a/a),\cm
            \Lambda=\gamma\rho.
\ear
  can be, alternatively, done as follows. It can be shown that all three
  models do not differ from one another from both mathematical and physical
  points of view.

  Let us denote
\bear
    f_1 = \alpha(\dot a/a)^2, \qquad
    f_2 = \beta(\ddot a/a),  \qquad f_3 =\gamma\rho.
\ear
  Then Einstein's equations for spatially flat space-time used in this work
  (\eqs (3) and (7)) can be written in the form
\bear
    f_1 + 0 f_2 - \frac{8 \pi G}{3} f_3 + \frac{1}{3}\Lambda \eql 0,
\\
    0 f_1 + f_2 + \frac{4 \pi G}{3}(1 + 3w)f_3 +\frac{1}{3}\Lambda \eql   0.
\ear
  This is a set of linear algebraic equations for $f_1$, $f_2$ and $f_3$
  with constant coefficients. If we add to this set any linear equation of
  the form
\bear
    a f_1 + b f_2 + c f_3 = d\Lambda
\ear
  with constant coefficients $a$, $b$, $c$, $d$, then the Einstein equations
  and this new equation make a closed set of three linear algebraic
  equations for $f_1$, $f_2$, $f_3$ . Its solution for any real $a$, $b$,
  $c$, $d$ can be written as
\bear
    f_1 = \frac{1}{\alpha}\Lambda, \cm f_2 =  \frac{1}{\beta}\Lambda,
    \cm     f_3 = \frac{1}{\gamma}\Lambda.
\ear

  A comparison of \eqs (A.2) and (A.6) clearly shows that they are equivalent
  to \eq (A.1). However, an important point is the arbitrariness of $a$, $b$,
  $c$ and $d$. Therefore, using one of the relations in \eq (A.1)
  automatically leads to one of the other two relations, and consequently
  there is no difference in the dynamic behaviour of these models.

\Acknow {One of the authors (SR) would like to express his gratitude to the
  authorities of IUCAA, Pune, CTS, IIT-Kharagpur and ICTP, Trieste for
  providing him Visiting Programmes under which a part of this work was
  carried out. XHM is partly supported by NSF 10675062 of China and is also
  grateful to ICTP, Trieste for providing him a Visiting Programme. We all
  are thankful to the referees for valuable suggestions which have enabled
  us to improve the manuscript substantially.}

\end{document}